# TMA-Grid: An open-source, zero-footprint web application for FAIR Tissue MicroArray De-arraying

Aaron Ge[1], Monjoy Saha[1], Máire A. Duggan[2], Petra Lenz[3], Mustapha Abubakar[1], Montserrat García-Closas[4], Jeya Balasubramanian[1], Jonas S. Almeida[1], Praphulla MS Bhawsar[1]

1 - Division of Cancer Epidemiology and Genetics, National Cancer Institute, National Institutes of Health, Maryland, USA
2 - Department of Pathology and Laboratory Medicine, Cumming School of Medicine, University of Calgary, Calgary, AB T2N 1N4, Canada
3 - Leidos Biomedical Research Inc, Frederick, United States
4 - The Cancer Epidemiology and Prevention Research Unit, The Institute of Cancer Research, London, SM2 5NG, United Kingdom.

## Abstract

**Background:**

Tissue Microarrays (TMAs) significantly increase analytical efficiency in histopathology and large-scale epidemiologic studies by allowing multiple tissue cores to be scanned on a single slide. The individual cores can be digitally extracted and then linked to metadata for analysis in a process known as de-arraying. However, TMAs often contain core misalignments and artifacts due to assembly errors, which can adversely affect the reliability of the extracted cores during the de-arraying process. Moreover, conventional approaches for TMA de-arraying rely on desktop solutions.Therefore, a robust yet flexible de-arraying method is crucial to account for these inaccuracies and ensure effective downstream analyses.

**Results:**

We developed TMA-Grid, an in-browser, zero-footprint, interactive web application for TMA de-arraying. This web application integrates a convolutional neural network for precise tissue segmentation and a grid estimation algorithm to match each identified core to its expected location. The application emphasizes interactivity, allowing users to easily adjust segmentation and gridding results. Operating entirely in the web-browser, TMA-Grid eliminates the need for downloads or installations and ensures data privacy. Adhering to FAIR principles (Findable, Accessible, Interoperable, and Reusable), the application and its components are designed for seamless integration into TMA research workflows.

**Conclusions:**
TMA-Grid provides a robust, user-friendly solution for TMA dearraying on the web. As an open, freely accessible platform, it lays the foundation for collaborative analyses of TMAs and similar histopathology imaging data.

# Introduction

Tissue Microarrays (TMAs) have revolutionized histopathology by allowing for the simultaneous scanning and multiplex analysis of potentially hundreds of tissue samples on a single slide (Kononen et al., 1998). Once digitally scanned, these TMAs can be processed to extract individual cores for analysis. This process, called de-arraying, is far more efficient than scanning each tissue sample individually, both in terms of time and resources required. However, distortions and errors during TMA manufacturing, assembly, sectioning, staining, and scanning often undermine their full potential. These issues include core misplacement, heat-induced deformation, fragmentation, and other problems (Nguyen et al., 2018). Accurate de-arraying requires precise identification of tissue cores and their alignment with the expected TMA grid to correctly associate the cores with corresponding case- and study-level metadata. Although several de-arraying methods have been proposed previously, a notable limitation of many of them lies in the segmentation step, where tissue is differentiated from the background. A common assumption made by these implementations is a bimodal distribution of the image's pixel intensity, where the two dominant peaks represent background and foreground. Methods based on this assumption range from manually setting a global threshold value (Vrolijk et al., 2003) to more sophisticated ones like Otsu's method (Wang et al., 2011). Pre-processing methods such as contrast enhancement and template matching (Dell'Anna et al., 2005), and post-processing using morphological operators have also been reported to improve segmentation results (Wang et al., 2011). However, this bimodal assumption often fails with images from new fluorescence devices due to complex backgrounds (Nguyen et al., 2018). In fluorescence imaging, background elements like dust and stains often appear as high-intensity peaks in the histogram, while tissue cores may have lower intensities. Consequently, using a high threshold may miss many cores, and a low threshold may include numerous outliers. Nguyen et al. address these limitations by applying a locally adaptive thresholding on an isotropic wavelet transform of the input TMA image (Nguyen et al., 2018); however, their implementation lacks interactivity to allow manual correction of cases where the algorithm fails.

Furthermore, most of the methods in the literature either have no publicly available implementations (Nguyen et al., 2018), involve non-trivial installation and setup (Schapiro et al., 2022), or presuppose the availability of powerful computing resources. Easy to use desktop solutions such as those provided by scanner vendors, typically come with significant deployment and licensing costs. These solutions often end up being data enclaves, making interoperability with other tools difficult to achieve. Open-source solutions like QuPath, while addressing concerns with deployment siloing, typically require the images to be locally available, posing governance and reproducibility issues. These issues are significant roadblocks to the widespread use of these tools by clinicians and pathologists. For this reason, there has been a movement in the scientific community towards FAIR research, which emphasizes that both data management and computational constructs should be aligned with principles of Findability, Accessibility, Interoperability, and Reusability (FAIR)(Wilkinson et al., 2016).

In line with recent efforts to move histopathological analysis into the web computing environment ([Bhawsar et al., 2023](#)) to promote such FAIRness, we developed TMA-Grid, an interactive, zero-footprint web application to address the aforementioned gaps in TMA de-arraying (see Availability for the URL to the fully functional fully working client-side application). This application was developed entirely in JavaScript and can be used on any device that supports a web browser, including commodity computers, smartphones, and tablets. TMA-Grid works directly with whole slide imaging (WSI) data generated by digitally scanning the TMA slide. Furthermore, using the ImageBox3 framework for web-based WSI traversal (Bhawsar et al. 2023), this client-side application can operate on the slide at multiple resolutions as required, thus alleviating the need for powerful hardware.

The design of TMA-Grid was continuously refined through repeated evaluation and feedback from a group of pathologists. Acknowledging the fact that an algorithmic approach to de-arraying cannot account for all potential edge cases, the web application allows users to interactively adjust the output at every step of the de-arraying process, drawing inspiration from QuPath's user-friendly design (Bankhead et al., 2017). The main advantage of our tool is its delivery of a powerful dearraying platform that requires no download or installation and is highly interactive, particularly in the way it allows users to correct segmentation and gridding results on TMAs that potentially cannot be de-arrayed automatically.

# Methods

Similar to previous literature, we split the de-arraying problem into two steps: 1) tissue segmentation, which involves extracting tissue cores from the slide background and removing any artifacts, and 2) gridding, which aligns the identified tissue cores in the best possible grid, taking into account missing or malpositioned cores. Below, we outline the preparation of the dataset used to train the pixel-wise segmentation model and how the segmentation result was used to detect cores from the image. We also describe how the identified cores were arranged in an optimal grid, with any missing or misplaced cores highlighted as such.

## Dataset Generation

For our dataset, we selected 119 TMA whole slide images from the Polish Breast Cancer Study (PBCS) (García-Closas et al., 2007) and 2 prostate cancer TMA whole slide images from Harvard Dataverse (Zhong et al., 2017). These images underwent various types of immunohistochemical staining, including BCL2, CD163, CK56, EGFR, HE, and HER2, and were downsampled to 512 x 512 pixels. This resolution was chosen to preserve essential details while enabling fast, in-browser segmentation. We manually annotated the centroids of each core in image pixels, using a uniform radius for all cores. This approach, while not accounting for manufacturing-related core irregularities, was intentional to train the model to predict where tissue should be, allowing TMA-Grid to correctly classify damaged samples as single entities.

To enhance model generalization, we used the Albumentations library (Buslaev et al., 2020) to create 20 augmented versions of each original image. Augmentation techniques included random flipping, rotation (up to ± 45 degrees), affine transformations, and the addition of Gaussian blur and fog. This process yielded 2420 augmented PNG images, which formed our training dataset. For each augmented image, we created corresponding 512 x 512 binary masks, separating circular cores from the background. The final training dataset consisted of these 2420 augmented images and their masks, with the downsampled whole slide images serving as input and the binary masks as the expected segmentation output. The 121 original images were reserved for testing and validation purposes, ensuring an unbiased evaluation of the model's performance.

## Training the Segmentation Model

Similarly to (Schapiro et al., 2022), we used TensorFlow to build a U-Net convolutional neural network for tissue segmentation as shown in Figure 1. This approach demonstrates superior performance over traditional tissue segmentation methods, particularly in cases of fragmented or dimmed cores. Unlike the standard U-Net architecture (Ronneberger et al., 2015), we replaced regular convolutions with separable convolutions in the convolutional blocks to improve model efficiency and reduce model size. Additionally, we introduced batch normalization and dropout layers to enhance model regularization and improve generalization performance. The model applies ReLU activation in both its contracting and expansive paths, along with max pooling for feature extraction and increased computational efficiency. Dropout layers are employed to achieve regularization. In the expansive path, the model combines features through upsampling and concatenation, producing a 512x512 binary segmentation mask as the output. Figure 2 shows an image segmented by the model. To address data imbalances, the model uses weighted binary cross-entropy loss.

The trained model was assessed on a held-out test set of 12 TMA whole slide images. Various metrics characterizing model performance were computed, including loss, Area Under the Curve (AUC), accuracy, precision, and recall, as presented in Table 1. The model itself is open-source and available for use from the same GitHub repository as TMA-Grid (see Availability).

## Core Detection

To detect core boundaries in the whole slide image, we started by processing the segmentation mask. First, we applied Otsu's thresholding (Otsu et al., 1979) to separate foreground from background by filtering out low-confidence areas, making it easier to distinguish objects in the image.

Next, we fixed small gaps or "holes" that could be a result of manufacturing errors or tissue breakage. We did this by comparing two versions of the image: one where the segmented areas were slightly expanded (a dilation operation) and one where the expanded regions were sharpened (an opening operation). This comparison helped us spot the small holes by generating an outline of all the objects of the image. We labeled these holes, measured their sizes, and decided what counted as a "small" hole based on a distribution of their sizes. Small holes were marked and filled in the original image, which prevented them from being mistaken for separate tissue areas. We then applied a distance transform, a method that helped highlight the main regions with tissue (the "sure foreground"). We labeled these main areas to create markers for the next step.

Using these markers, we applied the watershed algorithm to identify exact boundaries of the tissue cores. Once the areas were separated, we measured the size of the circular tissue areas and filtered out those that were too small or too large based on user-defined criteria.

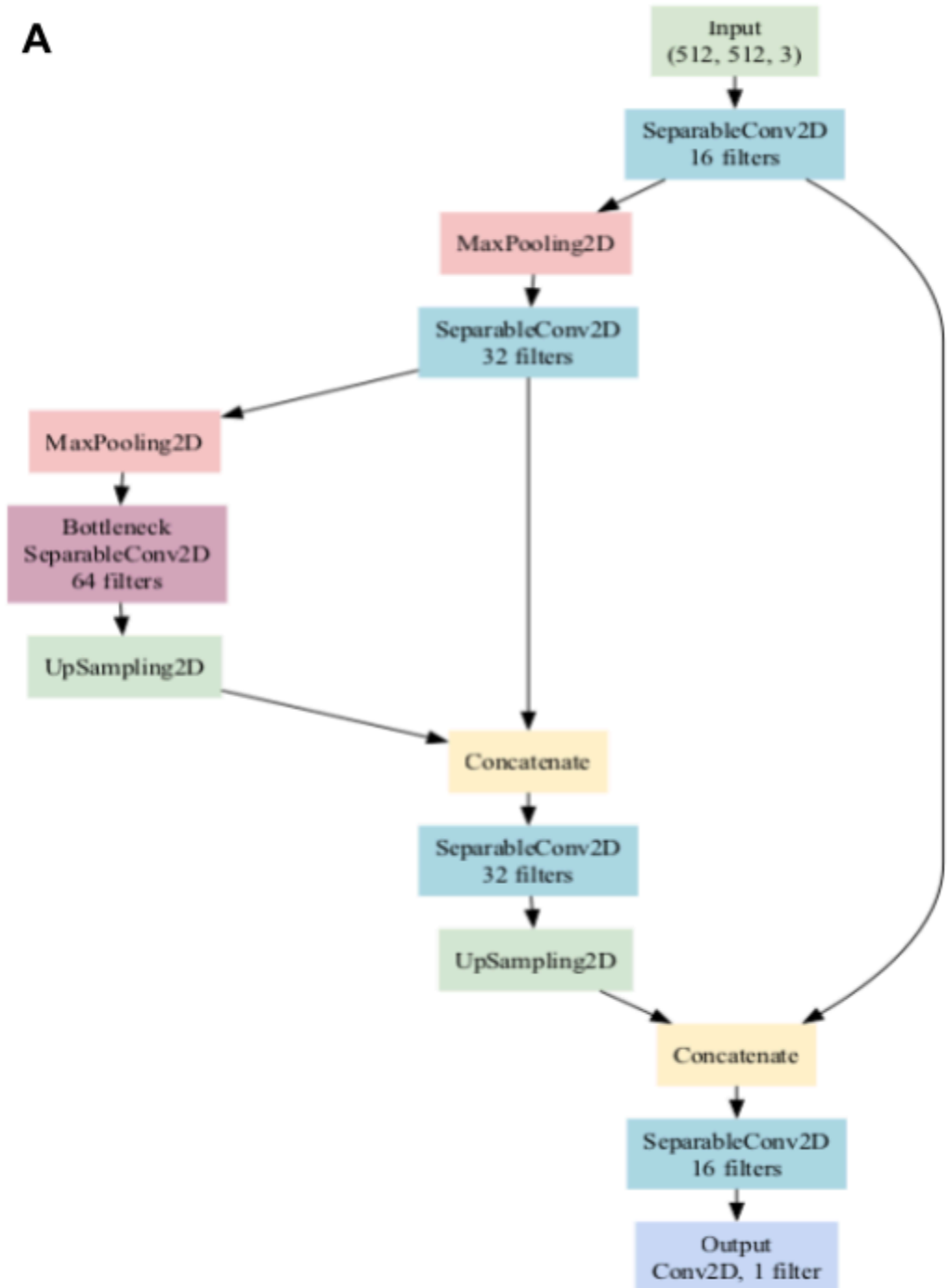

**Figure 1 - A)** Segmentation Model Architecture: The U-Net model is composed of a contracting path composed of separable convolutional blocks with increasing filter sizes, followed by max-pooling layers. The expansive path mirrors the contracting path but applies upsampling followed by concatenations with corresponding contracting path layers. Finally, a convolutional layer with a single filter generates the segmentation output. Dropout regularization is applied after each layer in the contracting path.

| Loss | AUC | Accuracy | Precision | Recall |
|---|---|---|---|---|
| 0.224 | 0.981 | 0.919 | 0.818 | 0.947 |

**Table 1** - Model performance evaluated on a held-out test set. The model demonstrates strong discriminative ability between tissue and background, with an AUC of 0.979.

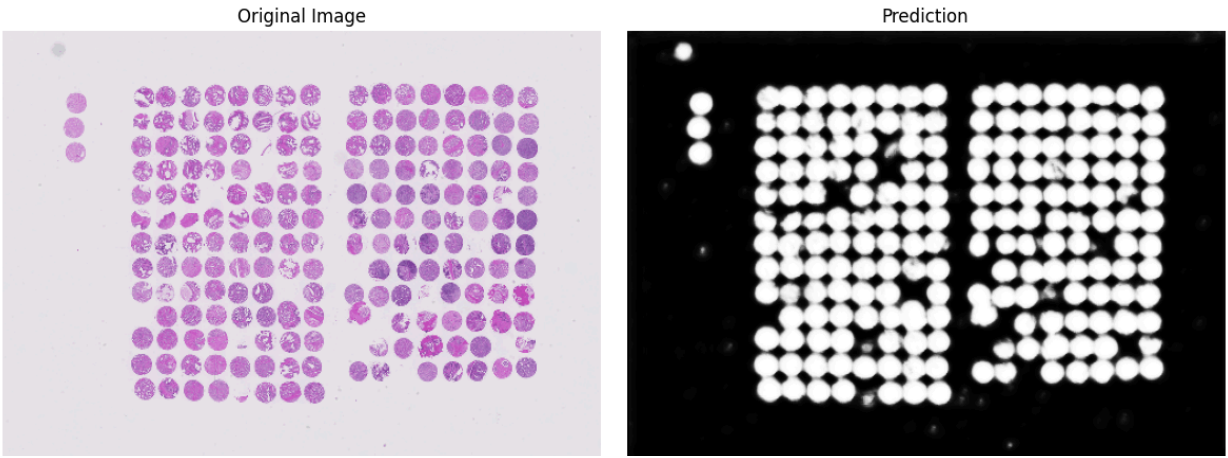

**Figure 2** - Segmentation Result: The original whole slide image is shown on the left and the mask predicted by the segmentation model is shown on the right.

# Gridding Algorithm

## 1) Delaunay Triangulation-Based Segment Generation

To impose a 2-dimensional grid on the identified centroids, we began by estimating row-wise connections between them in the form of segments connecting two or more cores. For this, we modified the Delaunay triangulation-based approach described by Wang et al. – triangulating points in a plane for edge generation, followed by filtering those edges based on length and angle. This process resulted in line segments defining a row, interspersed with isolated points (Fig. 2a).

The modification we made is in the criteria for filtering lengths and angles. While the earlier work filtered out all segments longer or shorter than 1.5 times the interquartile range (IQR), we argued that only segments longer than 1.5 times the IQR need to be excluded. We found that removing shorter segments often inadvertently filtered out connected points in a row, especially when cores were closely spaced. The result of the length-based filtering is shown in Fig. 2b.

For angle-based filtering, Wang et al. employed an unsupervised k-means clustering, categorizing angles into five groups based on their rotation [-90, -45, 0, 45, 90 degrees] and then selecting the cluster closest to 0 degrees. While effective for grids with minor deformities, this method struggles with significantly misaligned TMAs. We found that adopting a simpler thresholding approach for filtering segments worked better even for poorly aligned TMAs. Our algorithm filtered out all edges outside a certain threshold of an image's rotation (Fig. 2c). We empirically determined that a threshold of 10 degrees worked best for most TMAs in our training set. It is worth noting that there could be cases where this simpler approach might fail, but those can be easily resolved via the interactive features built into TMA-Grid.

An additional refinement in our algorithm is an edge cleaning step, which ensures that each point connects to at most two edges. In instances where a point was linked to more than two edges, our algorithm selected only the closest connected points in each direction in terms of Euclidean distance. This step was crucial for maintaining the integrity of the rows formed by the segments.

To summarize, the segment generation step organizes the TMA core centroids into a set of segmented points, i.e., those connected by a segment to one or more cores, and isolated points.

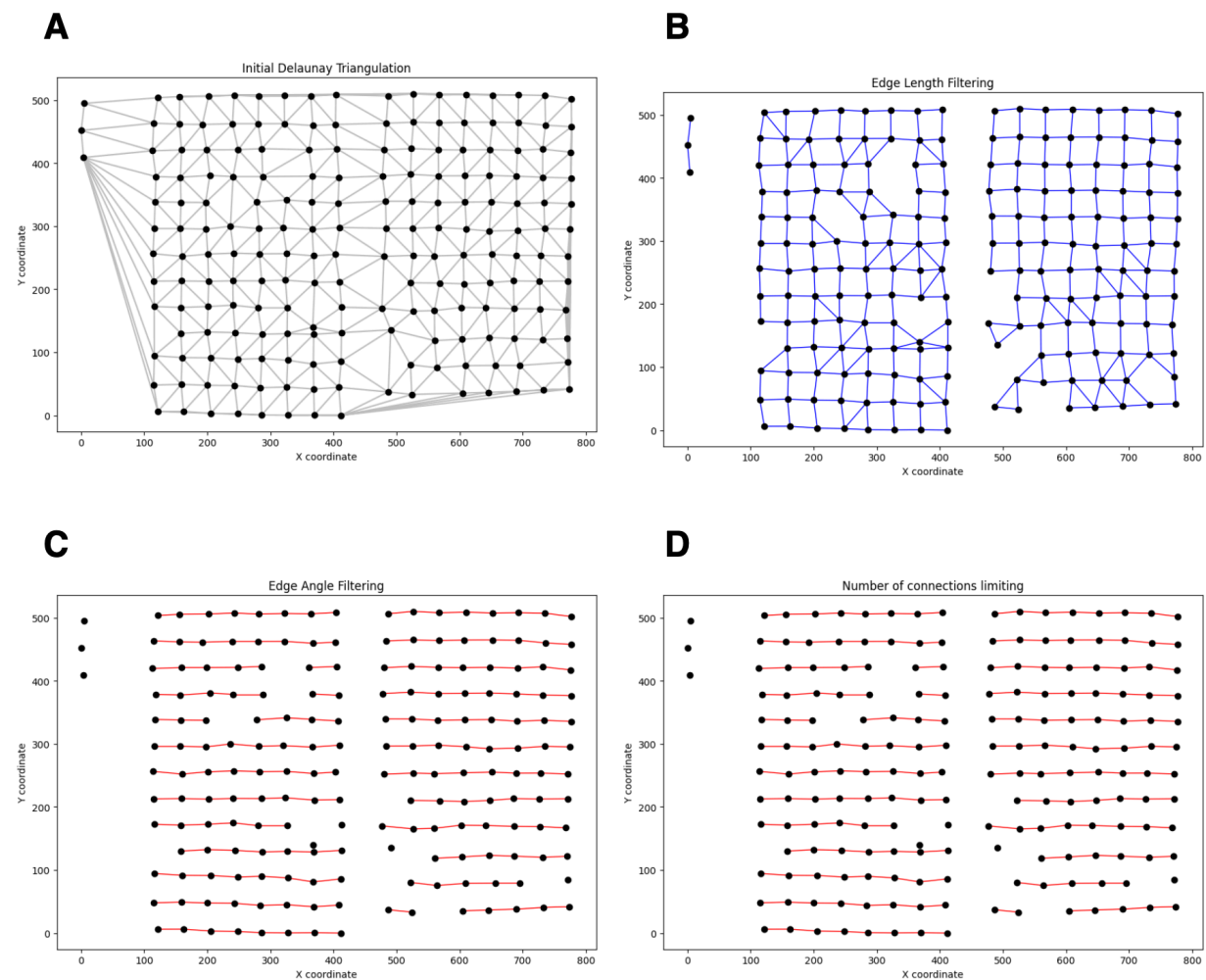

**Figure 2** - Results of Delaunay Triangulation-based Segment Generation: Starting with Delaunay triangulation to generate edges (2A), we filter edges based on length (2B) and angle (2C), distilling them into rows. Finally, we ensure that each point is, at most, connected to two points, by picking their closest

neighbor in the left and right direction (2D). The resulting segments define rows and include isolated points, those with no segments connecting them to another point.

## 2) Traveling Algorithm

The traveling algorithm described here iteratively identifies complete rows using the points and segments derived in the previous step. This method was also built upon Wang et al.'s traveling algorithm, with some key modifications. To begin with, each point was assigned a start point and an endpoint. For each point, we set either the point to its immediate right on the same segment as its endpoint, or itself if none satisfies those conditions. For an isolated point, the start and endpoints were both set to be itself. Note that image rotation might mean that the coordinates in the image are not representative of the arrangement of the actual TMA and can potentially cause issues in precise row assignment. TMA-Grid accounted for this by adjusting all coordinates for image rotation. We describe how the rotation is estimated and adjusted for in a later section.

During each iteration, the algorithm identified the start of a row by picking the point with the smallest value for the horizontal coordinate. For each point that is added to the row, the algorithm searches for another whose start point was the same as the endpoint of the added point. If such a point was found, it is added to the row and removed from the overall pool. If no point was found, the algorithm instead looked for points within a circular area with a radius of $r \times d$ around the endpoint of the current point, where $r$ is a user-specified parameter and $d$ is the median distance between all the points. By default, r is set to 0.6. This approach contrasts with the original algorithm (Wang et al., 2011), which identified suitable vectors by projecting a sector from the current point in a specific user-set direction. We found that detecting points within a circular region rather than a specific sector worked better in identifying misaligned cores. If multiple points were found within that radius, the closest core in terms of Euclidean distance was chosen, and the row generation continued.

If an appropriate point cannot be found within the search radius, the algorithm introduces an "imaginary" point to ensure continuity of the row. These imaginary or missing points are commonplace in TMA design: quite often, the microarray grid is not completely filled, with many spots showing no cores even in the middle of a row or column. Our traveling algorithm incorporated this in the computation of the row itself.

The end point for this imaginary core was set to be $[x + d \times cos(\theta), y + d \times sin(\theta)]$, where $(x, y)$ denoted the coordinates of the endpoint of the last point in the row, and $\theta$ denoted the rotation of the image. The algorithm would continue to create imaginary points until a real one was found or until the last point in the row reached the edge of the image, as defined by the width of the image subtracted by a user-defined stopping distance.

Once the edge was reached, the points were sorted by their horizontal coordinates. This sorting step is crucial for TMAs with rows that have several cores clustered compactly in a region. After sorting, the row was then completed, and the next iteration began. This continued until all points had been assigned to a row. The generated rows were then sorted again based on the vertical coordinates of their first point, after rotating each of the first points around the center of the image by -θ to account for image rotation.

Finally, in the case that some rows had points missing at the start, imaginary points were imputed to ensure that the start of each row was vertically aligned with the others as closely as possible. These imaginary points were generated backward, starting from the first real point in the row, using the equation $[\,x - d \times cos(\theta),\, y - d \times sin(\theta)\,]$, where $(x, y)$ were the coordinates of the first point in the row and $\theta$ was the rotation of the image.

## 3) Post-processing

Detecting sectors, which are sub-grids within a tissue microarray (TMA) that facilitate sample organization, is another necessary step for accurate row and column assignment. During row generation, columns may be filled with imaginary cores if the traveling algorithm repeatedly fails to find points within the defined radius. However, it is difficult to determine whether the absence of cores indicates a potential sector boundary or if the cores were significantly misaligned during construction. Some sector boundaries may even contain real cores, making sector identification more complex than simply finding empty columns. After running the traveling algorithm, the results were refined to handle specific elements, such as core sectors and marker cores (guides for pathologists in TMAs). First, misaligned cores and marker cores were identified by comparing each core's position to the median position of their respective columns. Cores deviating by more than one radius from the median were flagged as misaligned, while those deviating by more than 1.5 radii were considered marker cores. These marker cores were excluded from row and column indexing because their irregular placement affected the grid arrangement. Next, indices were reassigned, placing each core into the column closest to it in terms of vertical alignment. This process used a weighted distance formula that penalized columns with fewer cores, preventing misaligned cores from forming new columns.

$$D = \frac{d}{log\,(C + 0.000001)}$$

In the equation, d is the difference between the horizontal coordinate of the core and the median horizontal coordinate of the column, and C is the number of cores in that column. After this step, it was significantly less likely that the resulting rows or columns predominantly composed of imaginary cores could be a result of misalignment. These can then be removed by simple thresholding. TMA-Grid used thresholds of 75% imaginary cores for rows and 80% imaginary cores for columns – only if the number of imaginary cores were below the threshold was the row or column added to the indexing. In the end, index reassignment was conducted to ensure that the row and column values assigned to each core were consecutive and unique. The process concluded with a final check for misaligned and marker cores, prominently highlighting any that might require manual adjustment in the web visualizer.

## 4) Determining Image Rotation

To determine the image's rotation, we iteratively apply the processing pipeline described above at multiple image rotation angles, starting from -10° to +10° in one-degree increments. This range balanced efficiency with the expectation that major image distortions were corrected pre-analysis. For each angle, we adjusted centroid coordinates accordingly and reran the segment generation and traveling algorithm. For each angle, we assessed three proxy metrics for alignment quality: the counts of unique rows, imaginary cores, and marker cores. The optimal rotation minimized all three. When several angles yielded comparable results, we selected the median angle as optimal, except in the case where one of the optimal angles is 0°, in which case it was chosen by default.

# Results

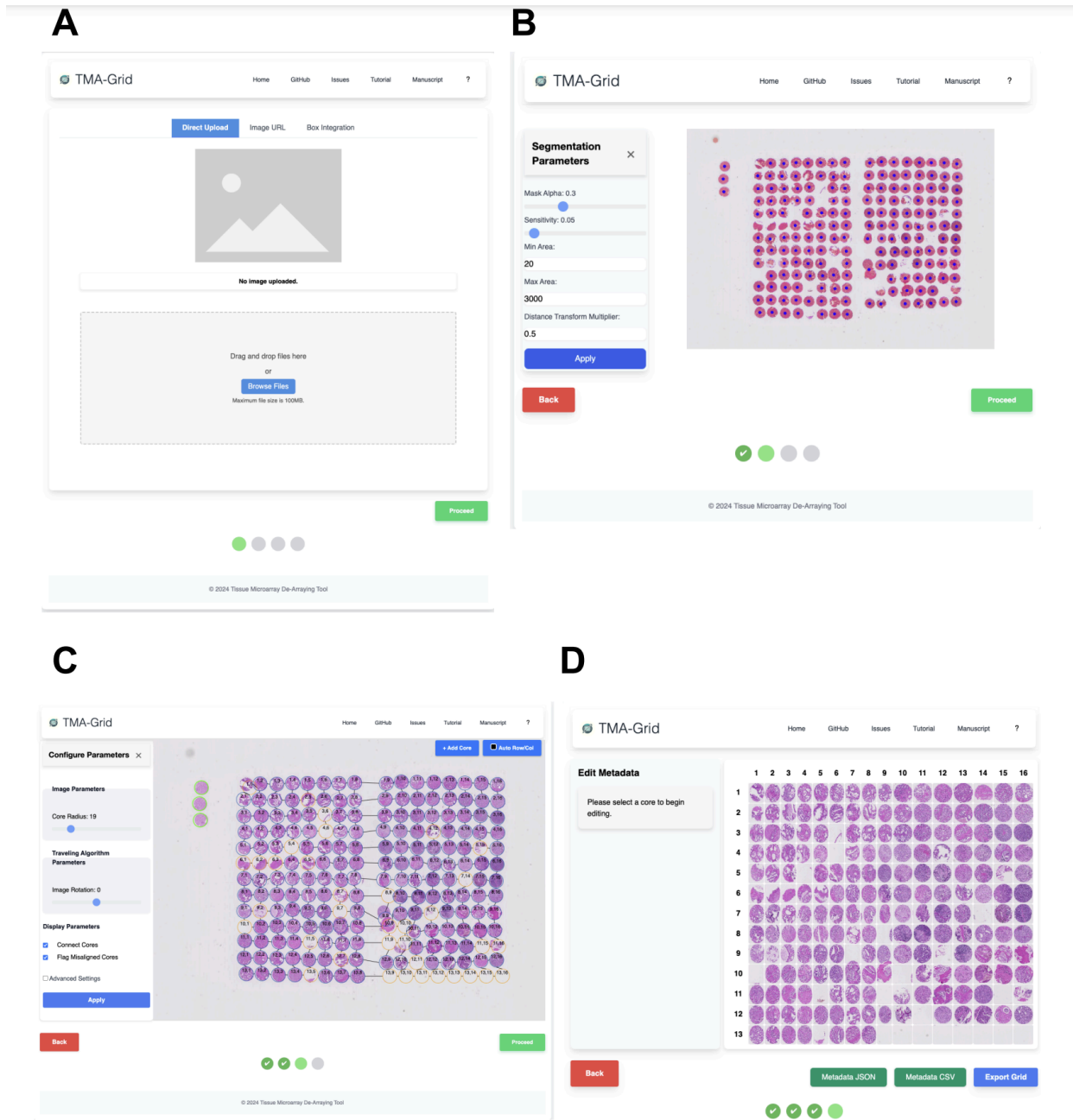

**Figure 3)**: The TMA-Grid web application **A) Data Input:** The initial upload interface. TMA-Grid allows WSI to be added from the local device, remote URLs, or via Box. The application uses the ImageBox3 library to retrieve regions of the image on demand. **B) Core Segmentation**: The segmentation result, with adjustable parameters such as the "sensitivity" threshold, which determines the cutoff probability for marking a region as tissue. Additionally, users can manually add or remove cores detected by the algorithm by clicking on the screen. **C) Grid Estimation**: Interactive whole slide viewer with an overlaid grid on top of the segmented cores and the ability to customize parameters. Core position, size, and row/column indexes can also be changed here, with each input triggering a reconstruction of the grid and providing real-time feedback on the operation's result. **D) The Virtual Grid:** An idealized virtual grid previewing the dearrayed cores. Core-level attributes can also be modified here, and new fields/annotations can be added to help in downstream validation or analysis. A table describing all the attributes per core can be generated as a JSON or a CSV file, along with the full-resolution versions of the dearrayed cores.

TMA-Grid is a versatile web application designed to streamline the analysis of whole slide images (WSIs) for tissue microarrays (TMAs). This tool offers an efficient workflow from data input to final grid generation, allowing users to manage and analyze TMA cores seamlessly. Figure 3 provides a detailed overview of the TMA-Grid interface and its key functionalities.

**Data Input**

TMA-Grid allows users to select images for analysis from various sources, including local devices, network systems, remote URLs, or Box accounts. Instead of uploading the entire image, TMA-Grid uses the ImageBox3 library to retrieve regions of the image on demand, optimizing computational efficiency. Once an image is selected, the application displays a preview for user verification (Figure 3A).

**Core Segmentation**

After selecting the image, users proceed to the tissue segmentation and core detection step. TMA-Grid generates a downsampled 512 x 512 version of the image for segmentation by a neural network. Users can adjust the "sensitivity" threshold, which determines the cutoff probability for marking a region as tissue (Figure 3B). For instance, a "sensitivity" threshold of 0.4 marks pixels with a model-predicted probability of 0.6 or greater as tissue. Users can fine-tune other parameters to optimize core detection and manually add or remove cores detected by the algorithm through an interactive interface.

**Grid Estimation**

Once segmentation is verified, users move on to the grid generation step, which involves detecting rows and columns of cores. This step provides extensive customization options for the gridding algorithm, including:

- Image Rotation: The tool computes an optimal rotation through trial-and-error, but users can change the values set by the algorithm.
- Radius Multiplier: Determines the search radius for disconnected cores.

- Stopping Distance Parameter: Guides the algorithm on when to conclude the current row and start a new one.

Users can edit core attributes such as size, row/column assignment, and add manual annotations for later review. Additionally, cores can be added, removed, or repositioned using click-and-drag operations. Each input triggers an automatic grid reconstruction, providing real-time feedback (Figure 3C).

**Virtual Grid**

In the final review step, TMA-Grid creates an idealized virtual grid of dearrayed cores, removing unnecessary background and arranging the cores in a perfectly spaced grid for preview. Users can edit core information and generate new fields for downstream validation or analysis. Once the dearraying result is satisfactory, a table describing all core attributes can be generated as a JSON or CSV file. Full-resolution versions of each core can be downloaded individually or collectively, maintaining fidelity through local PNG downloads (Figure 3D).

A comprehensive demonstration of TMA-Grid's functionalities is available on YouTube at this link.

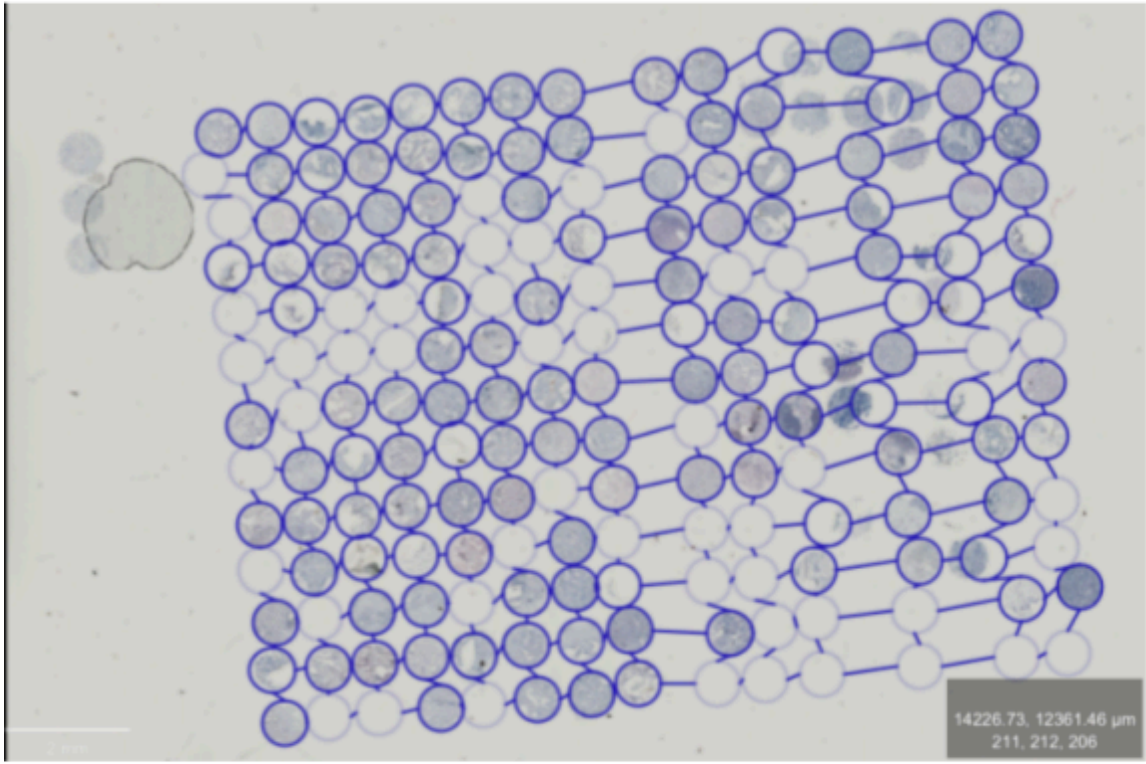

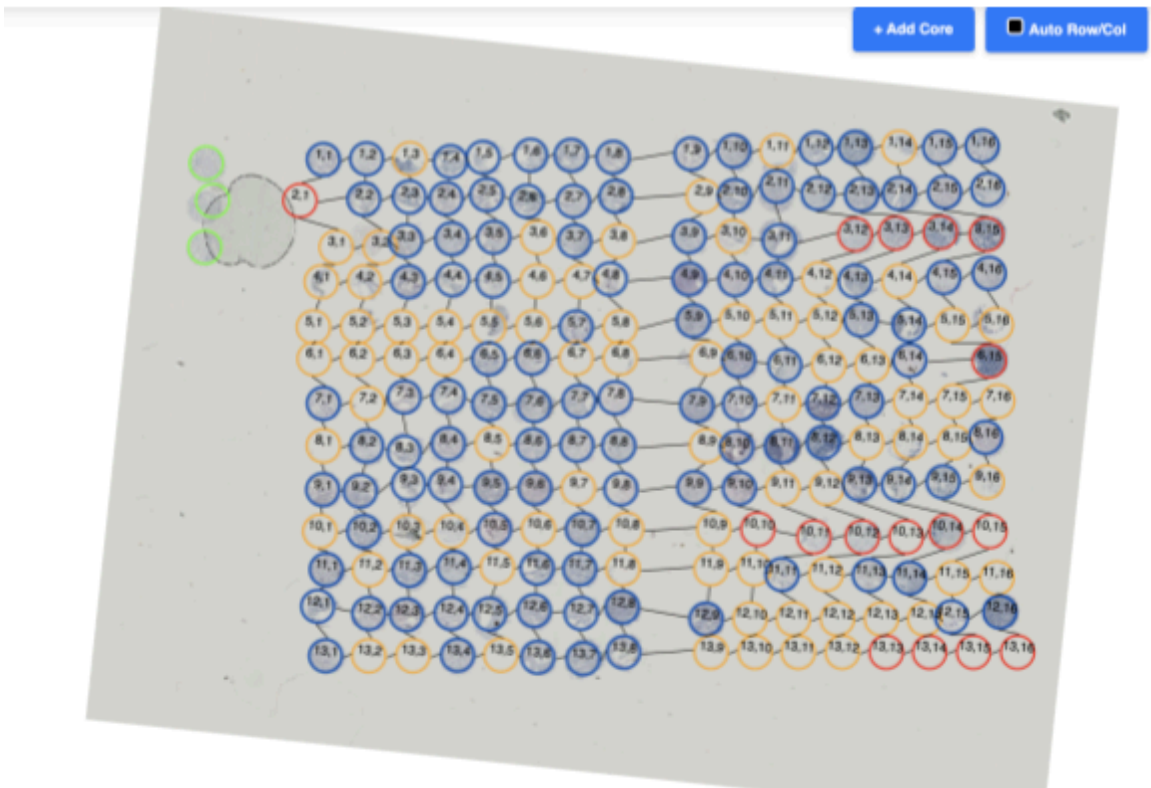

**Figure 4)** A comparison of the default de-arraying results using a) QuPath, and b) TMA-Grid on the same TMA whole slide image. For QuPath, it was necessary to manually apply values for the core diameter and row and column indexes, whereas TMA-Grid achieves comparatively better results by automatically optimizing all hyperparameters.

# Discussion

The tool described here is one of the first instances of a web-based, freely accessible, interactive analytics pipeline for WSI data. To the best of our knowledge, it is the only platform that allows zero-footprint dearraying of tissue microarray (TMA) cores without needing to upload the data to a specialized environment or have it be locally available. This is a marked improvement from traditional digital pathology workflows that require server-side patch extraction and prediction on whole slide images, resulting in multiple copies of the same data and invariably leading to governance and provenance concerns. In addition, there is a cost associated with the server even when these systems are not in use. This cost escalates when the tool or the analytics need to be distributed outside the clinic or institution, essentially prohibiting collaboration.

TMA-Grid addresses these issues by employing a client-side-only model. There is no specialized environment that exists independently of the user's machine. Instead, the environment is assembled at runtime when the user visits the application via their browser. In this model, the data does not need to be moved anywhere. The dearraying code operates on the data where it is located, regardless of whether it is local or on a remote data store. With a login-based mechanism, as employed with Box in TMA-Grid, the tool acts on behalf of the user, meaning that only users previously authorized to access particular data can use the tool on that data. There are no copies that need to be maintained and governed, and there is no cost, either for self-use or for distributed deployment.

It should be noted that Box is only meant to be a proxy here for a cloud service. The tool leverages Box's OAuth offering to allow users to operate upon their images directly from the app, completely under their own governance. Any storage service that offers similar functionalities for remote access can be similarly integrated with TMA-Grid. Just about any storage service that offers a pathway for delegated governance can be similarly integrated with TMA-Grid with minimal effort. This includes most, if not all, cloud-based storage providers, as well as most PACS software.

Despite running the tissue segmentation model on the user's device, the model takes on average 500 milliseconds to distinguish tissue from background on a multi-gigabyte image. However, one limitation of client-side operation is at the final step, when the high-resolution dearrayed cores need to be exported. This process involves significant computation and takes on average 10 minutes to complete on a commodity computer, depending on whether the image is local or on a remote service.

As part of the dearraying, it is necessary to establish a bijective mapping between the dearrayed cores and the TMAMap, the metadata file generated during TMA construction that contains associated clinical information for each core. The virtual grid functionality as described above could be especially useful here. TMA-Grid facilitates the association by generating a JSON/CSV table that maps each core to a row and column index, which can then be manually merged with the TMAMap by comparing these indexes. Although the tool could be modified to perform this mapping itself, we found that TMAMaps vary widely in how they are organized. It ends up being easier to manually create and validate the mapping via the TMA-Grid interface.

# Conclusion

Our work introduces a FAIR, interactive web application for de-arraying TMA whole slides, employing an AI model for tissue segmentation and a Delaunay triangulation algorithm for grid estimation. This approach is shown to be robust even against significantly misaligned TMAs. Furthermore, the availability of interactive parameter tuning options leads to better results over algorithmic pipelines. TMA-Grid proves the viability of using the web as a computational environment for digital pathology analytics. It sets the stage for creating virtual TMAs for instance, facilitating the assembly of individual cores from varied sources into comprehensive virtual arrays. Such flexible sample organization could streamline research processes, enabling easy data sharing and reduced resource use, and leading to more effective, portable and reproducible histopathological analytics.

# Funding

This work was funded by the National Cancer Institute (NCI) Intramural Research Program (DCEG/Episphere #10901).